\documentclass[referee]{raa}           
\usepackage{graphicx,times}
\usepackage{natbib}
\usepackage{amssymb,amsmath}
\usepackage{rotating, graphicx}
\usepackage{xfrac}
\bibpunct{(}{)}{;}{a}{}{,}

\usepackage[a4paper=true,dvipdfm=true,pagebackref=true]{hyperref}
\hypersetup{pdftitle = MRSSimulation, pdfauthor = Bo Zhang, pdfsubject= The subject, pdfkeywords = keyword1 keyword2 keyword3} 
\hypersetup{colorlinks = true, linkcolor = green, anchorcolor = red, citecolor = blue, filecolor = red, pagecolor = red, urlcolor = red}

\newcommand{\teff}{${T}_{\rm eff}$}
\newcommand{\logg}{$\log{g}$}
\newcommand{\feh}{$\mathrm{[Fe/H]}$}

\newcommand{\mh}{$\mathrm{[M/H]}$}

\newcommand{\am}{$\mathrm{[\alpha/M]}$}
\newcommand{\cm}{$\mathrm{[C/M]}$}
\newcommand{\nm}{$\mathrm{[N/M]}$}

\newcommand{\kms}{${\rm km\,s}^{-1}$}

\begin{document}

\title{Exploring the spectral \textit{information content} in the LAMOST medium-resolution survey (MRS)}
\volnopage{ {\bf 20XX} Vol.\ {\bf X} No. {\bf XX}, 000--000}
\setcounter{page}{1}

\author{Bo Zhang\inst{1,2}, Chao Liu\inst{1,2}, Chun-Qian Li\inst{1,2}, Li-Cai Deng\inst{1}, Tai-Sheng Yan\inst{1}, Jian-Rong Shi\inst{1}}

\institute{
Key Laboratory of Optical Astronomy, National Astronomical Observatories, Chinese Academy of Sciences, Beijing 100101,  People's Republic of China; {\it bozhang@nao.cas.cn}\\
\and
University of Chinese Academy of Sciences, Beijing 100049, People's Republic of China\\
\vs \no
{\small Received 20XX Month Day; accepted 20XX Month Day}
}

\abstract{
	Although high-resolution stellar spectra allow us to derive precise stellar labels (effective temperature, metallicity, surface gravity, elemental abundances, etc.) based on resolved atomic lines and molecular bands, low-resolution spectra are proved to be competitive in determining many stellar labels at comparable precision. 
    It is useful to consider the spectral \textit{information content} when assessing the capability of a stellar spectrum in deriving precise stellar labels.
    In this work, we quantify the \textit{information content} brought by the LAMOST-II medium-resolution spectroscopic survey (MRS) using the gradient spectra as well as the coefficients-of-dependence (CODs). In general, the wavelength coverage of the MRS well constrains the stellar labels but the sensitivities of different stellar labels vary with spectral types and metallicity of the stars of interest. This, as a consequence, affects the performance of the stellar label determination from the MRS spectra. Applying the SLAM method to the synthetic spectra which mimic the MRS data, we find that the precision of the fundamental stellar parameters \teff, \logg\ and \mh\ are better when combining both the blue and red bands of the MRS. This is especially important for warm stars since the H$\alpha$ line located in the red part plays a more important role in determining the effective temperature for warm stars. With blue and red parts together, we are able to reach similar performance to the low-resolution spectra except for warm stars. However, at $\mathrm{[M/H]}\sim-2.0$\,dex, the uncertainties of fundamental stellar labels estimated from MRS are substantially larger than that from low-resolution spectra. We also tested the uncertainties of \teff, \logg\ and \mh\ from MRS data induced from the radial velocity mismatch and find that a mismatch of about 1\,km\ s$^{-1}$, which is typical for LAMOST MRS data, would not significantly affect the stellar label estimates.
    At last, reference \textit{precision limits} are calculated using synthetic gradient spectra, according to which we expect abundances of at least 17 elements to be measured precisely from MRS spectra.
}

\authorrunning{Zhang et al. }            
\titlerunning{LAMOST MRS Information Content}  
\maketitle

%
\section{Introduction}           
\label{sect:intro}

A huge amount of spectral data with good quality are obtained through large-scale spectroscopic surveys, such as the SEGUE \citep{2009AJ....137.4377Y}, RAVE \citep{2006AJ....132.1645S}, Gaia-ESO \citep{2012Msngr.147...25G}, GALAH \citep{2015MNRAS.449.2604D}, APOGEE \citep{2017AJ....154...94M} and LAMOST \citep{2012RAA....12..735D}.
On the one hand, it has brought us new insights into the formation and evolution of the Galaxy \citep{2016ARA&A..54..529B}.
On the other hand, it challenges the spectral modeling.
Consequently, machine-learning approaches, \cite[e.g.,][]{2015ApJ...808...16N,2019ApJ...879...69T,2019MNRAS.483.3255L,2019arXiv190808677Z} are widely applied in the field to provide precise stellar labels (fundamental stellar parameters \teff, \logg\ and elemental abundances [X/H] and etc., hereafter we call them stellar labels following \cite{2015ApJ...808...16N}) at industrial scales \citep[cf. ][and the references therein]{2019ARA&A..57..571J}.

As argued by \citet{2017ApJ...843...32T}, the precision of stellar labels derived from spectra is determined by the \textit{information content} quantified by gradients imbedded in the spectra, which could be characterized mainly by spectral resolution ($R$), wavelength coverage and signal-to-noise (S/N) ratio and also depends on spectral types.
Traditionally, low-resolution spectra ($R<5,000$) are suitable for spectral classification, deriving fundamental stellar parameters and a few elemental abundances.
For example, \teff\ can be easily derived from Balmer lines.
Medium-resolution spectra ($5,000<R<10,000$) are generally sufficient for analysis in many studies of stars, and high-resolution spectra ($R>10,000$) are needed for very detailed analysis and determination of very reliable abundances \citep{2014dapb.book.....N}.

Although \citet{2017ApJ...843...32T} concludes that low-resolution spectra remain competitive for their low cost--performance ratio, the role of high-resolution spectra is the cornerstone in spectral analysis \citep{2019ARA&A..57..571J} while prices such as long exposure time and limited wavelength coverage have to be paid to obtain them.
Since their stellar labels can be confidently determined, they offer a "standard / reference" for other observations \cite[e.g.,][]{2008AJ....136.2070A,2014A&A...564A.133J,2015A&A...582A..81J,2015A&A...582A..49H,2016A&A...591A.118S} and are even "transferred" to low-resolution spectra \citep{2017ApJ...841...40H,2017ApJ...836....5H,2017ApJ...849L...9T,2019arXiv190808677Z,2019arXiv190809727X}.
Besides, the abundant resolved atomic lines and molecular features in high-resolution spectra also help to derive accurate radial velocity, micro-turbulence and rotation velocity of stars, as well as the identification of spectroscopic binary systems.

After finishing its first five-year low-resolution survey (LRS) \cite[$3900\mathrm{\AA}<\lambda<9000 \mathrm{\AA}$, $R\sim1800$, cf. ][]{2012RAA....12.1197C,2012RAA....12..735D,2012RAA....12..723Z,2015RAA....15.1095L} since September 2012, LAMOST (the Large Sky Area Multi-Object Fiber Spectroscopic Telescope) proceeds to conduct a new five-year medium-resolution survey (MRS, Liu et al. in prep.) since September 2018.
The MRS operates at $4950\mathrm{\AA}<\lambda<5350 \mathrm{\AA}$ (B band) and $6300\mathrm{\AA}<\lambda<6800 \mathrm{\AA}$ (R band) with spectral resolution of $R\sim7500$.
The MRS aims for several scientific goals, e.g., Galactic archaeology, stellar physics, star formation, Galactic nebulae, etc, most of which require precise stellar labels based on the MRS spectra.

Taking advantage of the high efficiency in acquiring spectra resulted from the 4,000 fibers on the focal plane, the MRS database will be quite attractive.
However, the wavelength coverage of the MRS is very limited.
For LRS, data-driven methods can derive precise stellar labels.
For example, \citet{2019arXiv190808677Z} derive \teff, \logg, \mh, \am, \cm, \nm\ at precision of $\sim$ 49 K, 0.10 dex, 0.037 dex,
0.026 dex, 0.058 dex, and 0.106 dex, respectively, for spectra with $g$-band signal-to-noise ratio $>100$.
Note that even the stars with multiple observations in the PASTEL \citep{2016A&A...591A.118S} catalog show a scatter of $\sim$50 K.
Therefore, it is worthwhile to think about how much more spectral information we can get from MRS compared to the LRS spectra.
In this paper, we try to quantify the \textit{information content} in the MRS spectra in two different ways, namely the gradient spectra and the coefficients of dependence (CODs), aiming to assess the performance of the MRS spectra in determining the stellar labels of F-, G- and K-type stars.
This paper is organized as following. In Section \ref{sec:infocontent}, we try to explore the spectral \textit{information content} in a general way. In Section \ref{sec:mrs}, we derive the precision of \teff, \logg\ and \mh\ from mock MRS spectra using the SLAM \cite[Stellar LAbel Machine, ][]{2019arXiv190808677Z}, a data-driven method, and also present a reference \textit{precision limit} of elemental abundances for MRS.
More discussions are shown in Section \ref{sec:discussion} and Section \ref{sec:conclusion} is the conclusion.


\section{Spectral Information Content}
\label{sec:infocontent}

The spectral \textit{information content} of a spectrum depends on spectral resolution, wavelength coverage and its stellar spectral type.
Quantifying the \textit{information content} in spectra given wavelength is important in traditional stellar spectral diagnostics, i.e., the Balmer lines can be used as proxies of \teff\ and almost independent of the overall metallicity \mh.
When choosing a wavelength range for a spectroscopic observation, one needs to think about how much information can be extracted from it.
However, this concept of spectral \textit{information content} was not systematically specified in previous works until \citet{2017ApJ...843...32T}.
Here we present two different methods to quantify the \textit{information content} of stellar spectra.
To demonstrate the quantification of \textit{information content} of stellar spectra for different types of stars, we select 8 sample stars including 4 spectral types (F-, G-, and K-dwarf and K-giant) and two metallicities ($\mathrm{[M/H]}=0.0$ dex and $-2.0$ dex) such that
\begin{enumerate}
	\item F-dwarf, \teff=7000 K, \logg=4.5 dex,
	\item G-dwarf, \teff=5800 K, \logg=4.5 dex,
	\item K-dwarf, \teff=4500 K, \logg=4.6 dex,
	\item K-giant, \teff=4500 K, \logg=1.8 dex.
\end{enumerate}
Then we generate mock spectra with \teff, \logg\ and \mh\ close to the parameters around each sample stars within $\pm1000$ K, $\pm0.25$ dex, and $\pm0.1$ dex, respectively.

\subsection{Gradient spectra -- a local measure}
\begin{figure}[!htbp]
   \centering
   \includegraphics[width=15cm, angle=0]{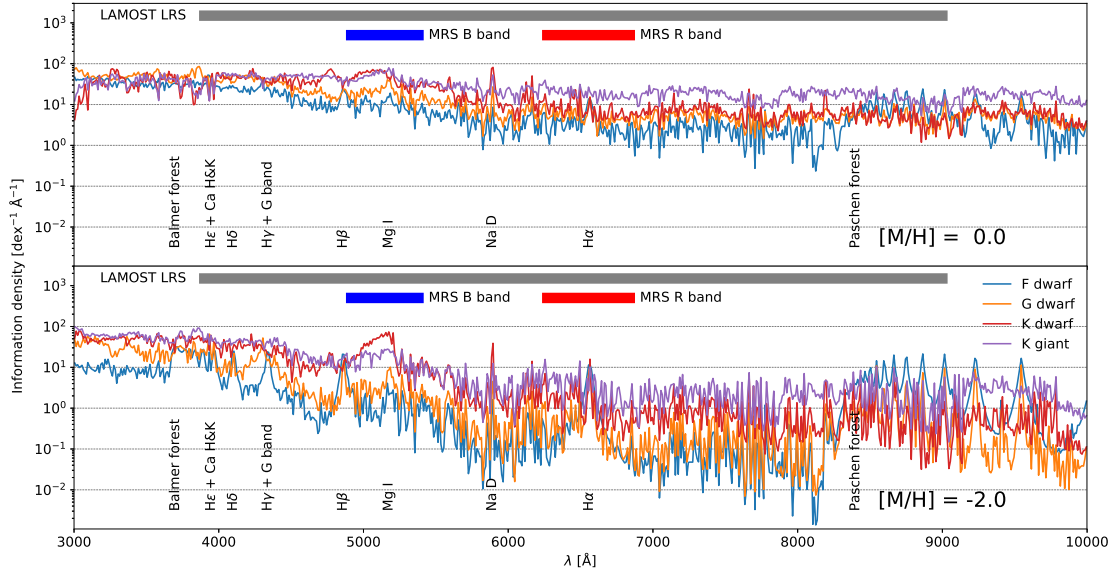}
   \caption{Each solid line represents the sum of gradient spectra. The blue/orange/red/purple line shows the results of F-, G- and K-dwarf and K-giant. The upper panel and lower panel shows the gradient spectra at $\mathrm{[M/H]} = 0$ and $-2$, respectively. Several well known spectral features are marked and the wavelength ranges of the LAMOST LRS and MRS are also shown in gray and blue/red bands.} 
   \label{fig:grads}
\end{figure}

The first way, as presented in \citet{2017ApJ...843...32T}, is to use gradient spectra to estimate the \textit{information content}.
Assuming there are $n$ stellar labels, $\vec{l}=(l_1,l_2,\cdots,l_i,\cdots,l_n)$, with the notation $\vec{l}+l_i=(l_1,l_2,\cdots,l_i+\Delta l_i,\cdots,l_n)$, the gradient of the spectrum on the $i$th stellar label is numerically calculated using
\begin{equation}\label{eq:grad}
    \dfrac{\partial}{\partial l_i} f(\vec{l},\lambda) = \dfrac{f(\vec{l}+l_i,\lambda)-f(\vec{l}, \lambda)}{\Delta l_i}.
\end{equation}
It measures the spectral response to variation of a given stellar label $l_i$. To quantify the total \textit{information content} relavent to fundamental stellar parameters, we plot the sum of gradient spectra, i.e., $\sum_{i} \Big\vert \dfrac{\partial}{\partial l_i} f(\vec{l},\lambda) \Big\vert$, following \citet{2017ApJ...843...32T} in Figure~\ref{fig:grads}, where the sum is over \teff, \logg\ and \mh.

In the upper / lower panel of Figure~\ref{fig:grads}, we show the sum of gradient spectra for sample spectra at $\mathrm{[M/H]}=0.0$ / $-2.0$ from 3,000 to 10,000 $\mathrm{\AA}$. The gradient spectra are evaluated using model spectra produced with ATLAS9 model\citep{2003IAUS..210P.A20C} at $R\sim300,000$ and binned to 10 $\mathrm{\AA}$ for visualization. A few well known spectral features and the wavelength spans of the LAMOST LRS and MRS are marked in the figure.

It is obvious that no matter at which metallicity, within this wavelength range, the blue part contains more information than the red part ($\sim1$ magnitude). In the lower metallicity case, the hydrogen features can be seen in the gradient spectra, such as the Balmer and Paschen features, especially in F-dwarf. From 3000 to 8000 $\mathrm{\AA}$, the gradient of F-dwarf decreases with wavelength more rapidly than cooler stars, which indicates that for warm stars more information of stellar labels is in the blue part. Beyond 8000 $\mathrm{\AA}$, as the Paschen lines arise, the gradient rises again.

On the other hand, late-type stars contain rich and significant metal lines and molecular bands. Although the blue part is more informative than the red, they are usually more luminous in the red part. Therefore, one has to compromise between the \textit{information content} and luminosity in practice to carry out a meaningful spectral observation.

\begin{figure}[!htbp]
   \centering
   \includegraphics[width=15cm, angle=0]{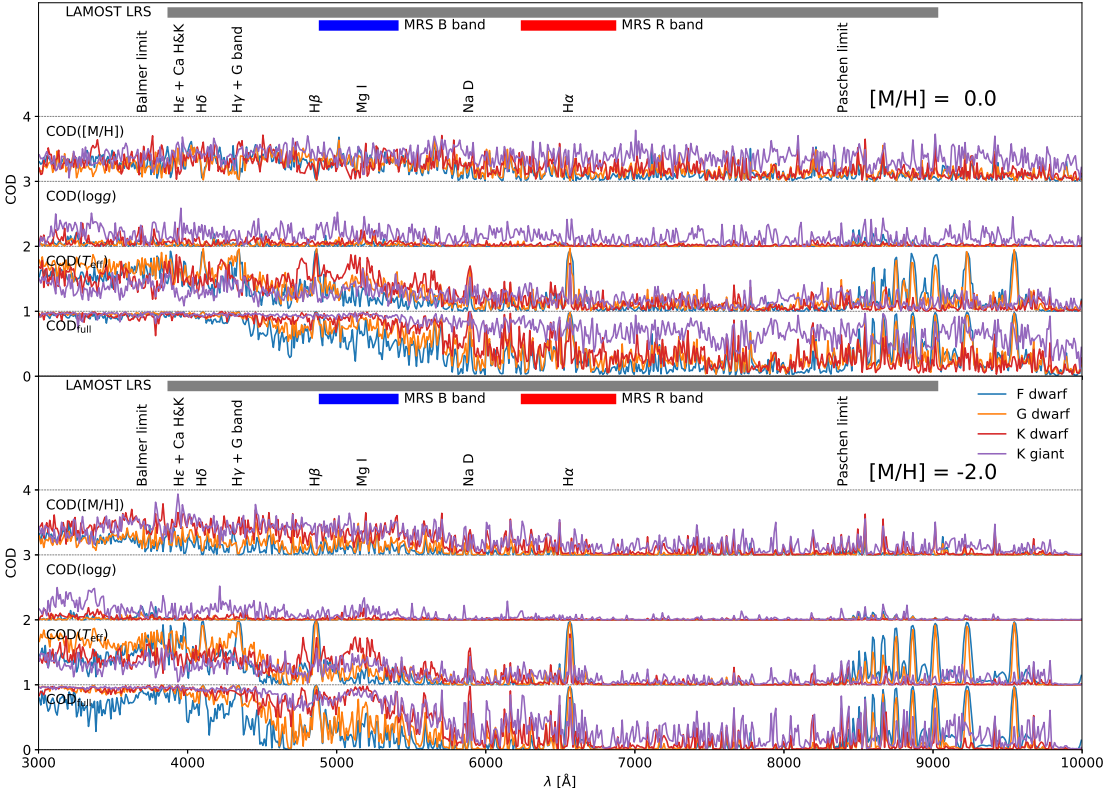}
   \caption{In the upper panel, each curve represents either the full COD or the COD for a specific stellar label. COD(\teff), COD(\logg) and COD(\mh) are shifted by a constant for visualization. The blue, orange, red and purple color traces the F-, G-, and K-dwarf and K-giant, respectively. Several well known spectral features are marked and the wavelength ranges of the LAMOST LRS and MRS are also shown in gray and blue / red bands. The lower panel shows the similar plot for $\mathrm{[M/H]} = -2$.} 
   \label{fig:cods}
\end{figure}

\subsection{CODs -- a global measure}
Secondly, \cite{2019arXiv190808677Z} used SLAM (Stellar LAbel Machine), a data-driven method, to evaluate the Coefficient of Dependence (COD) which quantifies the global spectral \textit{information content}. The basic idea is to measure the \textit{percentage of variance explained} (PVE) of spectral flux at each wavelength by regression. \textit{The full COD indicates the total spectral information content for determining all stellar labels, and the CODs for a single stellar label quantifies the spectral information content for that stellar label.}  
The advantages of CODs include that they can be evaluated from observed spectra with known stellar labels rather than synthetic spectra and CODs have unity scales.
We refer the readers to Appendix \ref{sec:ve} for the demonstration of how the PVE could be used to quantify the \textit{information content} in noisy data.

Here we briefly explain how to evaluate the full COD and the CODs for single stellar labels.
We define
\begin{equation}
\begin{split}
COD_{\rm full} = \sum_i COD(l_i) = PVE_{\rm full},
\end{split}
\end{equation}
where PVE$\rm _{full}$ is the variance explained when trained using all stellar labels, and $l_i$ denotes the $i$th stellar label.
As SLAM always produces a regression model that is close to \textit{ideal} by adopting adaptive model complexities for each pixel, we can assume that the COD$_{\rm full}$ is a simple sum of the contribution from each stellar label. To evaluate the COD of each stellar label separately, we do a leave-one-label-out training for each label. Let W$(l_i)$ be the relative contribution of $l_i$, PVE$(-l_i)$ be the PVE after excluding $l_i$ in training, from the leave-one-label-out training process we have the following linear equations
\begin{equation}
\left\{
\begin{array}{ccccccc}
       & +W(l_2) & +W(l_3) & +\cdots & +W(l_n) &= & PVE(-l_1)\\
W(l_1) &         & +W(l_3) & +\cdots & +W(l_n) &= & PVE(-l_2)\\
\vdots & \vdots  & \vdots  &  \ddots & \vdots  &= & \vdots\\
W(l_1) & +W(l_2) & +W(l_3) & +\cdots &         &= & PVE(-l_n)\\
\end{array}
\right. .
\end{equation}
Obviously, we have
\begin{equation}
    \sum_i W(l_i) = \frac{\sum_j PVE(-l_j)}{n-1}
\end{equation}
hence
\begin{equation}
    W(l_i) = \frac{\sum_j PVE(-l_j)}{n-1}-PVE(-l_i).
\end{equation}
The CODs for each stellar label can be derived via
\begin{equation}
\begin{split}
COD(l_i) & = COD_{\rm full} \times \dfrac{W(l_i)}{\sum_j W(l_j)} \\
& = PVE_{\rm full} \times \left(1- \dfrac{(n-1)PVE(-l_i)}{\sum_j PVE(-l_j)}\right) .
\end{split}
\end{equation}
\textit{They indicate the relevant fractions of spectral information content for determining each stellar label at a specific wavelength.}
Compared to gradient spectra, CODs have advantages including that they can measure the global sensitivity of the flux against the variance of stellar labels and can be directly evaluated from observed spectra as shown in \cite{2019arXiv190808677Z}. 
Interestingly, \cite{2019arXiv190808677Z} found that CODs are highly consistent with our traditional spectroscopic experience. For instance, the Balmer lines are good measures of \teff\ and almost independent of \mh, and the Mg I triplet at $5175~\mathrm{\AA}$ is a good proxy of \logg\ compared to other spectral features in $3900\mathrm{\AA}<\lambda<5800\mathrm{\AA}$.

\subsection{The information content of F-, G- and K-type stars in optical spectra}
We are able to evaluate the COD$_{\rm full}$ and the CODs of each stellar label for sample stars used in the previous subsection and display them in Figure~\ref{fig:cods}.

In the upper panel of Figure~\ref{fig:cods}, we show the CODs for stars with solar metallicity. For better visualization, we shift COD(\teff), COD(\logg) and COD(\mh) by a constant 1, 2 and 3, respectively. It is obvious that the COD$_{\rm full}$, which quantifies the total \textit{information content}, decreases with wavelength in the range from 3000 to 8000 $\mathrm{\AA}$. And cool stars have a higher \textit{information content} than warm stars at almost all wavelength, which is consistent with the gradient result.

In general, COD(\teff) traces the hydrogen features and metal lines while COD(\mh) traces the metal lines and molecular bands. The molecular bands are not significant since the effective temperature of the K-type stars in the test is not sufficiently low. The COD(\logg) remains low value except for K-type giant stars, meaning that it is relatively easy to determine \logg\ for K giant stars.

In the lower panel, we show similar results for low metallicity (${\mathrm{[M/H]=-2}}$) stars. The major difference is that all CODs are lower than those at solar metallicity.
The COD(\teff) strongly follows the hydrogen features and COD(\logg) almost vanishes at the red band. 
It is noted that, for metal-poor stars, the COD(\mh) is only significant in the blue band ($\lambda<6000 \mathrm{\AA}$). There are a few wavelength ranges where COD(\mh) is high, including $\lambda\sim3900~\mathrm{\AA}$, which is mostly contributed by Ca K and H lines. However, for metal-poor F- and G-type stars, the overall COD(\mh) in the MRS blue band is not prominent.

The MRS blue (B) / red (R) band is originally designed for the observations of Mg I triplet / ${\rm H}\alpha$. The figure shows that, for warm stars, the MRS R band has more information on \teff\ than the B band, while for cool stars the B band is more informative. In general, our results are consistent with the analysis of gradient spectra \citep{2017ApJ...843...32T} and also consistent with the traditional methods of measuring fundamental stellar parameters summarized in \citet{2019ARA&A..57..571J}.



\section{The expected Precision of Stellar labels from MRS}
\label{sec:mrs}
Empirically, we expect that the abundances of many elements could be determined with MRS which has $R\sim7,500$.
However, the precision of elemental abundance estimates highly relies on the precision of fundamental stellar labels.
In this section, we utilize SLAM \citep{2019arXiv190808677Z}, a data-driven method, to assess the performance of MRS spectra on stellar labels of F-, G- and K-type stars, and also derive reference \textit{precision limits} for many elemental abundances with gradient spectra.

\begin{figure}[!htbp]
  \centering
  \includegraphics[width=15cm, angle=0]{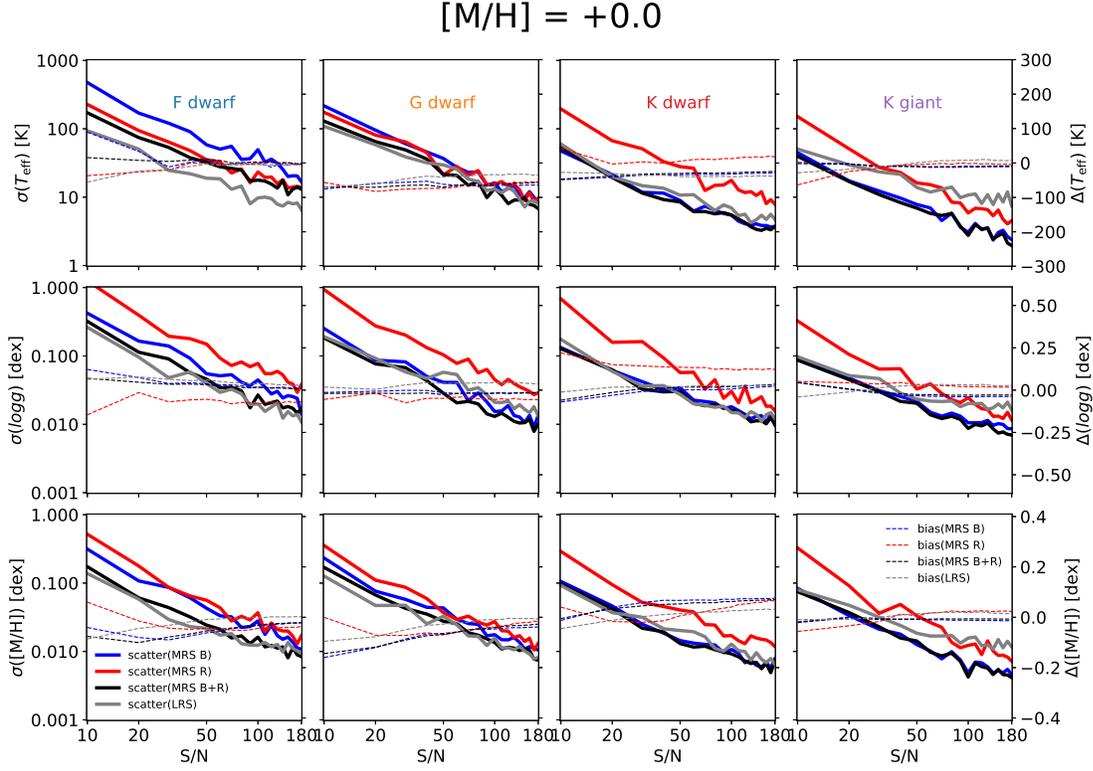}
  \caption{In each panel, the thick solid lines represent the scatter-S/N relation for sample stars with \mh$=0.0$. The blue/red/ black lines are calculated using MRS B band only / MRS R band only / both MRS B and R band spectra, respectively. The gray thick line is the result of the same test with LRS spectra. The dashed lines with corresponding colors are the bias. The first, second and third rows of the figure show the results for \teff, \logg\, and \mh, respectively. Each column represents one type of sample stars, which is marked in the top panels. Note that we use $\sigma$ to represent random error and use $\Delta$ to represent bias. In each panel, the left vertical axis denotes random uncertainties, and the right one denotes the bias.} 
  \label{fig:sigma0}
\end{figure}

\begin{figure}[!htbp]
  \includegraphics[width=15cm, angle=0]{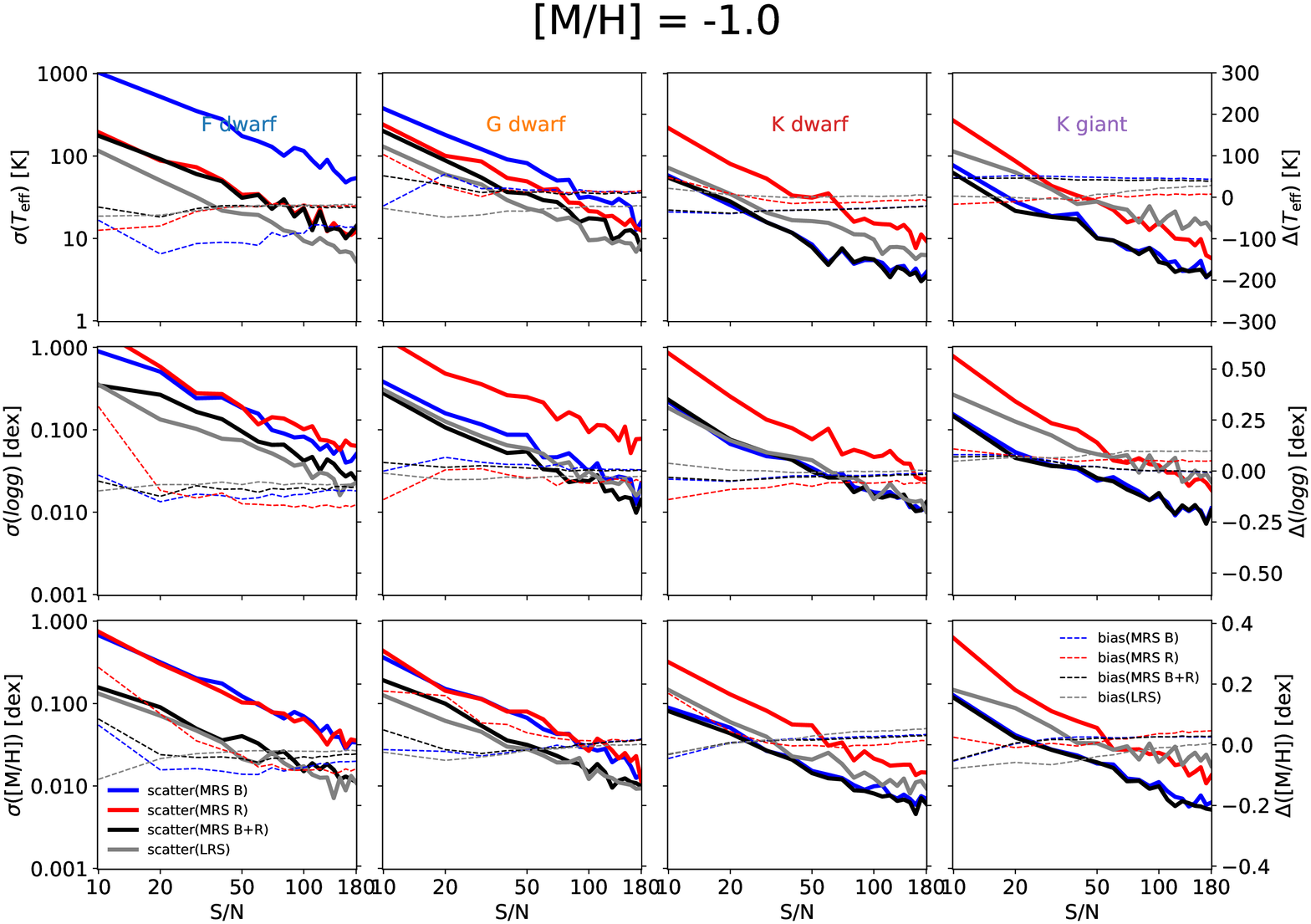}
  \caption{Similar to Figure~\ref{fig:sigma0} but for ${\mathrm{[M/H]=-1}}$.} 
  \label{fig:sigma1}
\end{figure}

\begin{figure}[!htbp]
  \centering
  \includegraphics[width=15cm, angle=0]{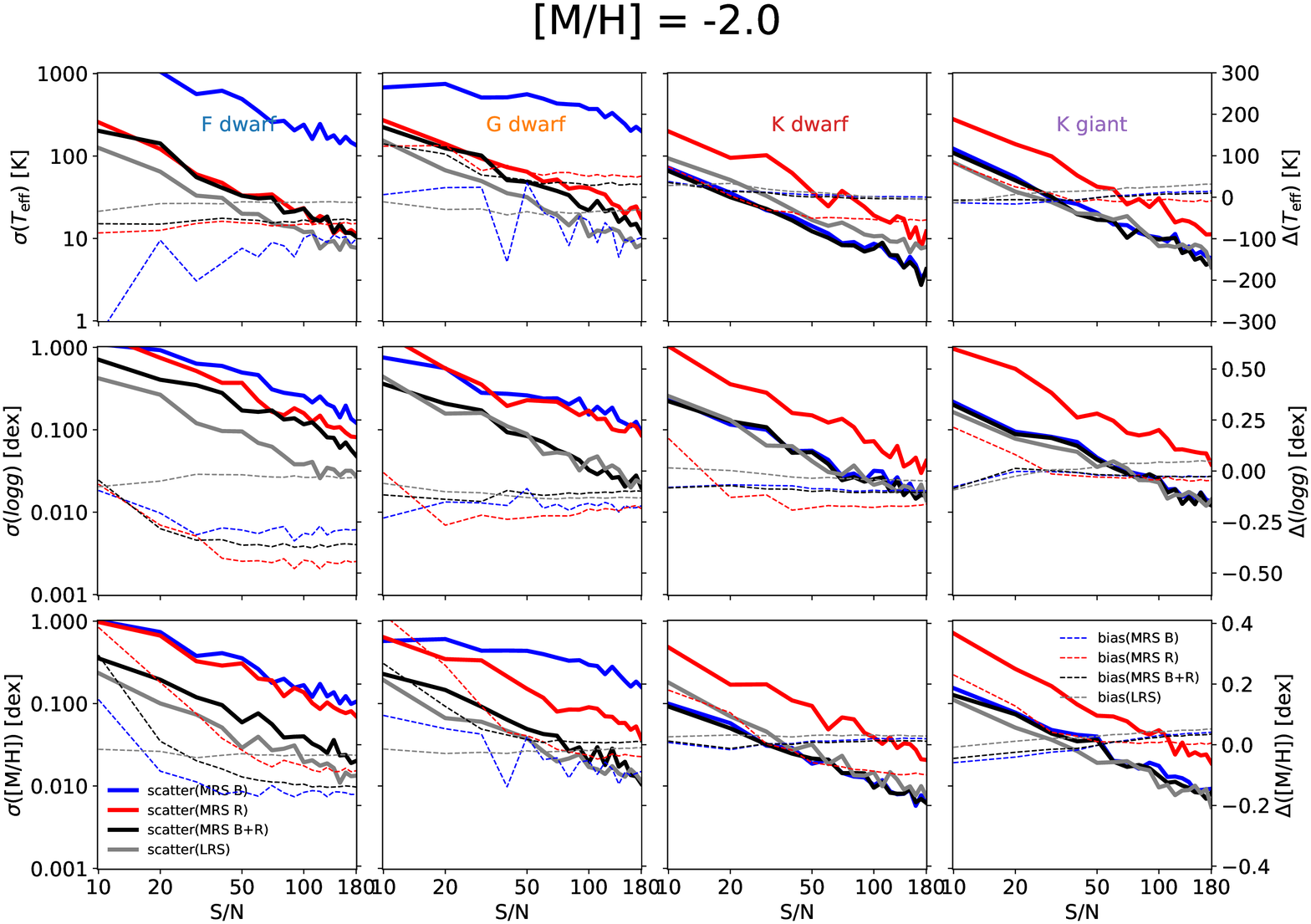}
  \caption{Similar to Figure~\ref{fig:sigma0} but for ${\mathrm{[M/H]=-2}}$.} 
  \label{fig:sigma2}
\end{figure}

\begin{figure}[!htbp]
    \centering
    \includegraphics[width=15cm, angle=0]{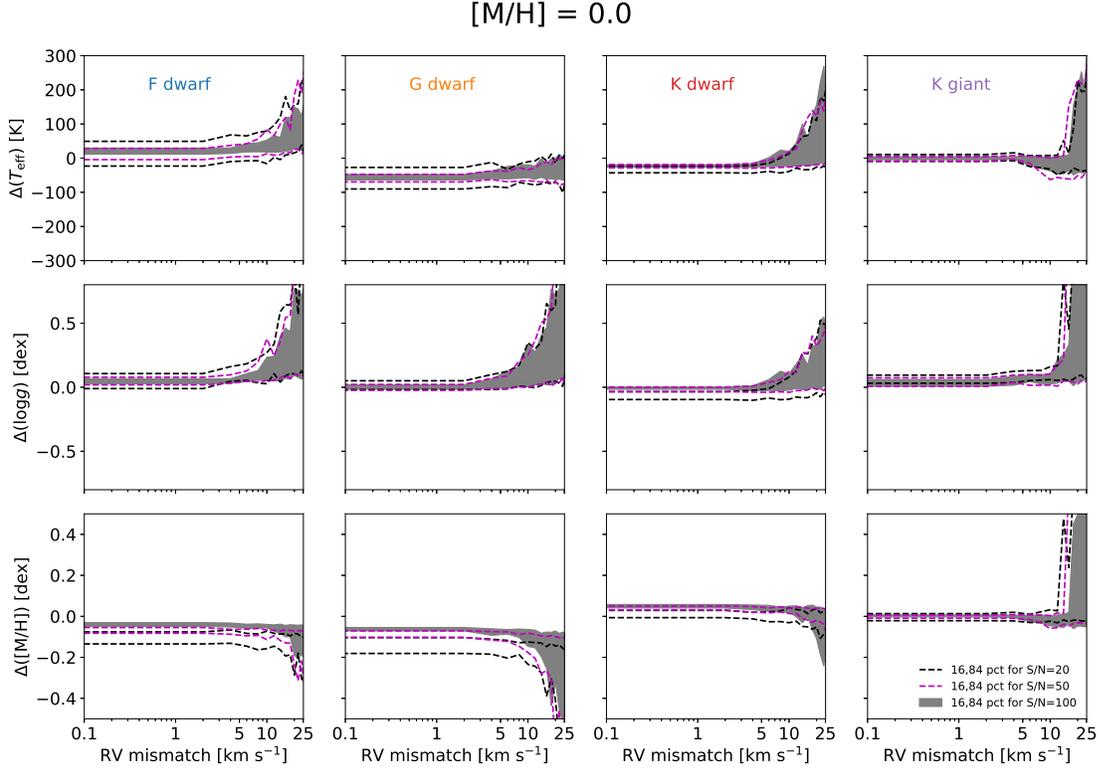}
    \caption{In each panel, the gray area, pink and black curve represents for the 16th and 84th percentiles of the replicated stellar labels compared to the true values. Each row shows the result for a specific stellar label and each column shows for one spectral type of stars.} 
    \label{fig:rv0}
\end{figure}

\begin{figure}[!htbp]
    \centering
    \includegraphics[width=15cm, angle=0]{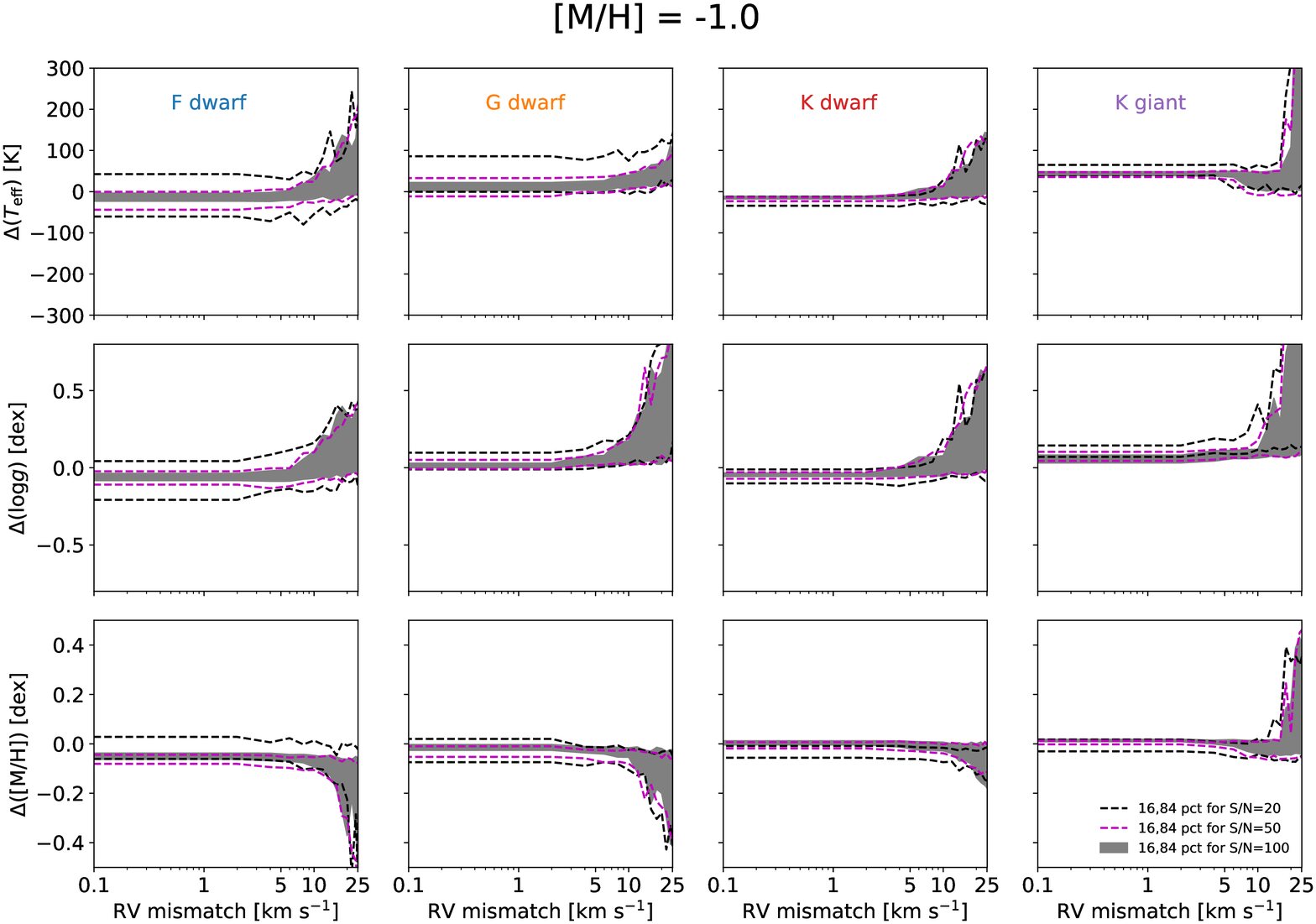}
    \caption{Similar to Figure~\ref{fig:rv0} but for \mh\ = -1.} 
    \label{fig:rv1}
\end{figure}  

\begin{figure}[!htbp]
    \centering
    \includegraphics[width=15cm, angle=0]{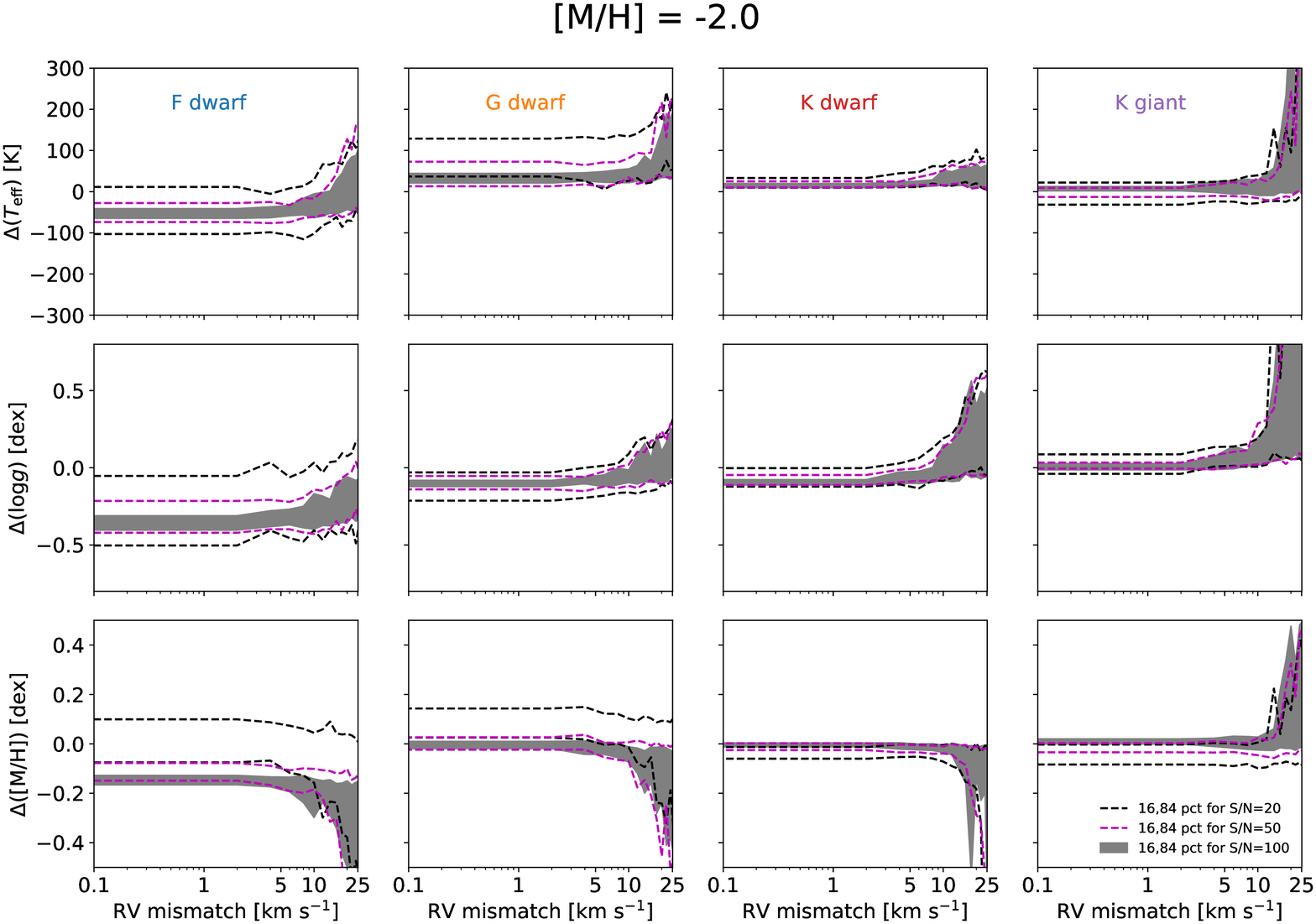}
    \caption{Similar to Figure~\ref{fig:rv0} but for \mh\ = -2.} 
    \label{fig:rv2}
\end{figure}

\subsection{Precision of fundamental stellar parameters \teff, \logg\ and \mh}
We use ATLAS9 to generate 6000 mock MRS spectra at $R\sim50,000$ and smoothed with a Gaussian kernel to degrade them to $R\sim7500$ with \teff\ between 3500 and 9000 K, \logg\ between 0 and 5 and \mh\ between $-$4 and 0.5.
To compare with LRS, we also generate another 6000 at $R\sim1800$ keeping other conditions the same. 
The MRS and LRS spectra are re-sampled to 0.2 $\mathrm{\AA}$ and 1.0 $\mathrm{\AA}$, respectively, to keep approximately the sampling steps at $\lambda/3R$.
To simulate the practical procedure of deriving fundamental stellar parameters using data-driven methods, noise is added to each spectrum so that the S/N is at 100. We trained SLAM separately with these two data sets.

To test the performance of MRS in deriving fundamental stellar parameters, again we study the four types of test stars used in Section \ref{sec:infocontent} but at three different metallicities (\mh $=$0, $-$1 and $-$2). Noise is added to each test star to mimic observed spectra at different S/N per pixel from 10 to 180.
Tests are repeated for 50 times at each S/N so that we are able to evaluate the bias and scatter for different combinations of test stellar spectra, metallicity, and S/N ratio. In Figure~\ref{fig:sigma0}, we show the results of our tests for $\mathrm{[M/H]}=0$.
In the top left panel, the thick solid lines represent the relationship between the scatter of \teff\ and S/N for F-dwarfs. The blue/red/black lines are calculated using MRS B band only / MRS R band only / both MRS B and R band spectra, respectively. The gray thick line is the result of a similar test but with LRS spectra. The dashed lines with corresponding colors are the bias of \teff\ estimates. The second and last rows are similar, but for \logg\ and \mh, respectively. From the second to the fourth column, we show the tests for G-, K-dwarfs, and K-giants. Figure~\ref{fig:sigma1} and \ref{fig:sigma2} are similar to Figure~\ref{fig:sigma0}, but for $\mathrm{[M/H]}=-1$ and $-2$, respectively.

In this series of figures, it is obvious that the MRS B band is more informative than the R band for K-type stars. The performance of any fundamental stellar parameter using the B band alone is close to that using the combination of B and R bands.
The reason is that for K-type stars, metal lines are abundant in the B band.

For G-type stars, the situation is similar to the K-type stars at solar metallicity, while it is quite different at $\mathrm{[M/H]}=-2$. As discussed in the previous section, the metal lines are relatively weak when metallicity is low.
Since B band is designed mainly for Mg I triplet and other metal lines and it does not cover any good \teff\ indicators such as hydrogen features, the B band lacks information of \teff.
As the \teff\ is the primary stellar parameter and may affect the performance of \logg\ and \mh\ estimates, all the three cannot be well determined using the B band only.

For F-dwarfs, the \teff\ derived from the B band alone is relatively uncertain compared to the R band., especially in the metal-poor case ($\mathrm{[M/H]}=-2$).
But the B band still has a precision of \mh\ comparable to the R band.
Hence combining both B and R band is important for F-dwarfs.
All three fundamental stellar parameters show larger uncertainties compared to G- and K-type stars.
It is reasonable because most metal lines are weak at this effective temperature.

Interestingly, although the LRS has a low resolution, it behaves quite robust across all metallicities and spectral types in these tests.
It is important to recall that the precision of stellar labels is determined by the total \textit{information content} in spectra with a given wavelength range.
Although the spectral resolution is low, covering from 3900 to 9000 $\mathrm{\AA}$ makes the LRS spectra contain similar (or even more) information in some situations than the MRS with narrower wavelength range.
Although we expect other elemental abundances from MRS to be more precise than the LRS, to determine the fundamental stellar parameters such as \teff, \logg\, and \mh, the LRS data is essentially more valuable.
Besides, combining with other spectroscopic, photometric, astrometric and asteroseismic data is also helpful to derive more precise stellar labels \citep{2019ARA&A..57..571J}.

\subsection{The influence of radial velocity mismatch}
It is necessary to correct radial velocity (RV) before deriving the stellar labels in most methods. Therefore, accurate RV is important in deriving precise stellar labels.
\cite{ 2019ApJS..244...27W} reported that the intrinsic precision of their RV measurements for spectra in MRS is able to achieve 1.36 \kms, 1.08 \kms\, and 0.91 \kms\ for the spectra at S/N ratio of 10, 20, 50, respectively.
However, the RV precision depends on stellar spectral types as well.
For example, K giants spectra contain abundant narrow metal lines, thus it is easy to obtain more precise RV than for A- and F-dwarf stars.
Xiong et al. (in prep.) develops a method to self-calibrate the RVs of a star using multiple epoch observations and analyzes the relation between RV precision and spectral types in more detail.
In their work, at S/N$\,\sim40$, the errors of RVs are generally under 0.7 \kms\ for almost all types of stars except B-type.
And Li et al. (in prep.) confirmed this RV difficulty for B-type stars.

For F-, G- and K-type stars, we present a simulation to estimate the response of stellar labels to the RV mismatch.
In Figure~\ref{fig:rv0}, \ref{fig:rv1} and \ref{fig:rv2}, we show that the deviation of stellar labels \teff, \logg\ and \mh\ against the RV mismatch at $\mathrm{[M/H]}=0$, $-1$ and $-2$, respectively.
The random RV mismatch used in this test is assumed to be Gaussian.
Tests are done in the same way as in the scatter-S/N test but an additional random error in RV is added to shift the test spectra.
And we also test the scatter-RV mismatch relation at three S/N ratios, i.e., 20, 50 and 100.

At almost all the three metallicities and all S/N ratios, a large RV mismatch introduces not only a scatter but also a significant bias of stellar labels.
The effect of erroneous RV tends to overestimate the \teff\ and \logg\ of all the four types of test stars (F-, G- and K-type dwarf and K giant) at all metallicity and all S/N ratios, while it tends to underestimate \mh\ of F-, G- and K-type dwarfs, except K-giants.

A reasonable explanation of this is that for spectra with relatively wide features (e.g., dwarfs), SLAM tends to predict best-matched spectra with shallower lines due to the RV mismatched spectra, so that the \teff\ and \logg\ are higher than the true values, while the \mh\ is of course lower.
For the K-giants, the different behavior of the bias of \mh\ is probably because most of the spectral lines are very narrow and deep.
However, within the reported precision of RV estimations, we do not see any significant increment of the scatter for any fundamental stellar parameter.

\subsection{Prospects of precise abundances of many elements from MRS}

\begin{sidewaysfigure}
  \vspace{15cm}
  \includegraphics[width=\textwidth, angle=0]{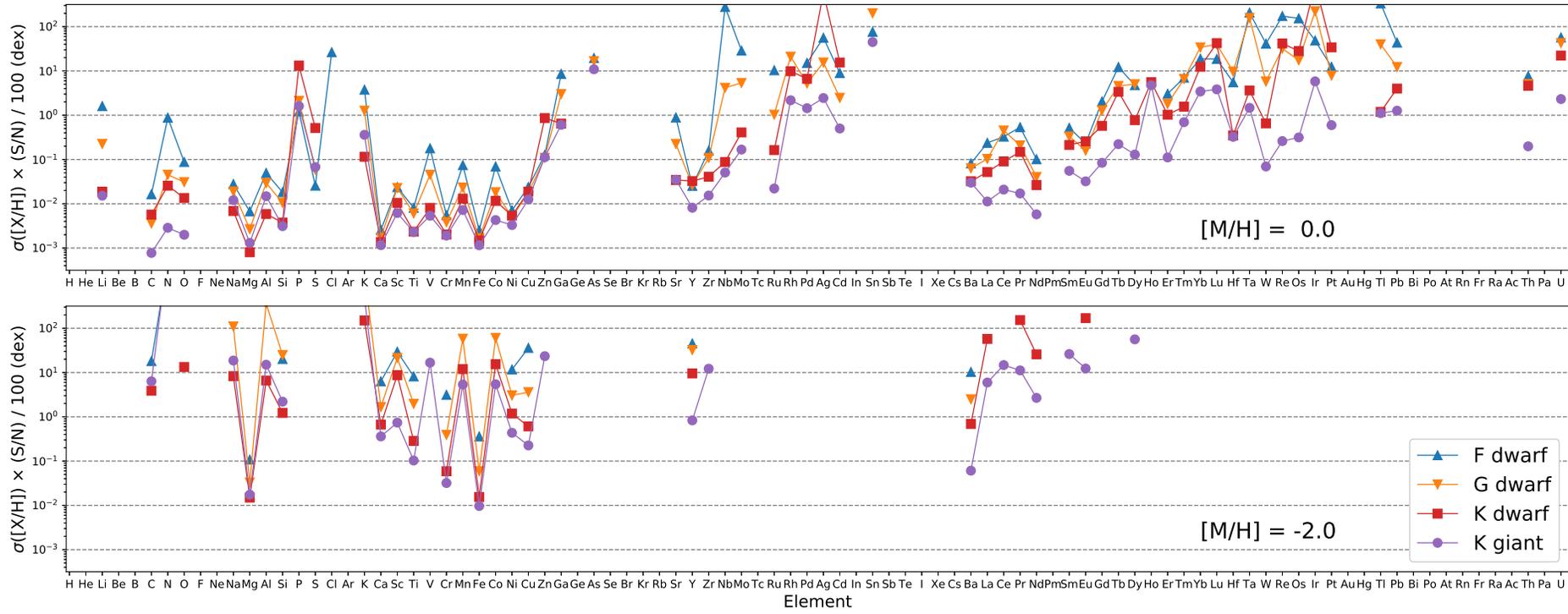}
  \caption{The upper / lower panel shows the \textit{precision limits} of elemental abundances for F-, G-, K-type dwarf and K-type giant stars in blue, orange, red and purple, respectively, at $\mathrm{[M/H]}=0$ / $-2$.} 
  \label{fig:sigma_atom}
\end{sidewaysfigure}

With synthetic gradient spectra, we are also able to predict the \textit{precision limits} of elemental abundances from MRS using a method similar to \cite{2017ApJ...849L...9T}.
The signal-to-noise ratio at $\lambda$ is $\mathrm{S/N}(\lambda) = \dfrac{f(\vec{l}, \lambda)}{\delta f(\vec{l}, \lambda)}$ by definition, where  $f(\vec{l}, \lambda)$ and $\delta f(\vec{l}, \lambda)$ are the normalized flux at $\lambda$ and its associated uncertainty.
Let $l_i$ represent $\mathrm{[X/H]}$, the elemental abundance under interest, the gradient spectrum on $l_i$ can be evaluated via Eq.~(\ref{eq:grad}).
Assuming that all pixels are uncorrelated with each other, the precision of the elemental abundance, $\sigma(l_i)$, is determined via
\begin{align}
    \dfrac{1}{\sigma(l_i)^2} &=\sum_\lambda \left( \dfrac{ \frac{\partial}{\partial l_i} f(\vec{l},\lambda)}{\delta f(\vec{l},\lambda)}\right)^2  \\
    &=\sum_\lambda \left( \dfrac{\mathrm{S/N}(\lambda) \times \frac{\partial}{\partial l_i} f(\vec{l},\lambda)}{f(\vec{l},\lambda)}\right)^2  .
\end{align}
We use ATLAS9 to generate normalized spectra at $R\sim 50,000$ for the sample stars defined in Section~\ref{sec:infocontent} (F-, G-, K-dwarf and K-giant) and degrade them to $R\sim7,500$ using a Gaussian smoothing.
In our test, the $\Delta\mathrm{[X/H]}$ is chosen to be 0.1 dex and $\mathrm{S/N}=100$ which is wavelength-independent.
Note that since these spectra are "born" on a normalized scale, we get around the \textit{pseudo-continuum normalization} step which contributes a large number of uncertainties in the reduction of observed spectra.
Therefore, our precision estimation is very optimistic and can be regarded as \textit{precision limit}.
We adopted solar abundance from \citet{1998SSRv...85..161G} and the \textit{precision limits} of $\sim 90$ elemental abundances are shown in the upper / lower panel of Figure~\ref{fig:sigma_atom} for $\mathrm{[M/H]=0}$ / $-2$.
The results for F-, G-, K-type dwarf and K-type giant stars are shown in blue, orange, red and purple, respectively.

In the $\mathrm{[M/H]=0}$ case (upper panel of Figure~\ref{fig:sigma_atom}, the \textit{precision limits}), there are many elements with $\sigma(\mathrm{[X/H]})\lesssim0.01$ dex.
These elements include C, N, O, Na, Mg, Al, Si, Ca, Sc, Ti, V, Cr, Mn, Fe, Co, Ni, and Y, and we expect precise abundances of at least these 17 elements come out from MRS.
In general, K-type giants provide the most precise elemental abundances among our 4 test stars.
When $\mathrm{[M/H]=-2}$, most elemental abundances become uncertain, except that Mg and Fe can still be measured precisely.
This is as expected because the MRS B band is designed for Mg I triplet at $\lambda\sim5175~\mathrm{\AA}$ and ion lines are abundant in the optical range.
Note that stars with enhanced elemental abundances are a special case.
For example, the carbon-enhanced metal-poor (CEMP) stars could have $\mathrm{[C/Fe]}>2$ \citep{2007ApJ...655..492A}, which means the carbon features are significant in spectra and remain detectable despite low $\mathrm{[M/H]}$.
And our \textit{precision limits} are S/N--dependent so that once $\mathrm{S/N}>100$ is achieved, the precision of elemental abundances could be better than shown and more elements can be measured.

\section{Discussion}
\label{sec:discussion}
\subsection{The precision of stellar labels}
In this work, we adopted SLAM as a representative of data-driven methods to simulate the process of deriving stellar labels.
The precision and bias of the SLAM-predicted stellar labels for the LRS are shown in Figure~\ref{fig:sigma0}, \ref{fig:sigma1} and \ref{fig:sigma2}.
At the high S/N ratio end, the precision of our stellar labels is very small.
For example, for the high S/N F-dwarf (S/N $\sim$ 100), the scatter of \teff\ is about 10 K.

On the one hand, this is due to the fact that our simulation is done ideally.
The random error of flux and the training error of SLAM are the only sources of the scatter of stellar labels.
In practice, the observed spectra may have bad pixels due to various reasons, and the pseudo-continuum normalization may introduce lots of uncertainties to the normalized spectra.
Therefore, the precision in practice will be worse than that in this work.
Typical scatters of \teff, \logg\ and \feh\ for LAMOST LRS spectra at $g$-band S/N higher than 100 are 50 K, 0.09 dex and 0.07 dex, respectively, as reported in \citep{2019arXiv190808677Z} using the 3900 to 5800 $\mathrm{\AA}$ spectra.

On the other hand, compared to the precision derived not with data-driven methods but with a synthetic model, such as \cite{2017ApJ...843...32T}, our scatters of stellar labels are quite similar to theirs.

\subsection{Caveat}
Several things are not taken into account in the test in this work. One of the most important is the influence of binary stars.
At $R\sim7,500$, a significant fraction of double-lined binary systems or even triple systems can be identified.
Considering the significant binary frequency in F-, G- and K-type stars \citep{2014ApJ...788L..37G}, it is important to identify whether the object is a single star or not before deriving stellar labels (Li. et al. in prep.).

\section{Conclusions}
\label{sec:conclusion}
As the LAMOST MRS is going on, it is important to assess the increase of the spectroscopic information compared to the previous LRS. We conclude our results below.
\begin{enumerate}
\item We explored the information quantification first, including the using gradient spectra and using CODs. As general-purpose tools, they are very helpful and valuable for astronomers working on stellar spectroscopy. It is easy to identify which wavelength window is more informative than others for a specific spectral type.
\item With these two tools, we can predict, in somehow, the performance of the MRS B and R band in deriving stellar labels. The LAMOST MRS B band is designed mainly for Mg I triplet and some other metal lines while the R band captures the ${\rm H}\alpha$ line. 
\item We utilized SLAM, a typical data-driven method, to simulate the process of deriving fundamental stellar parameters from MRS data. It is consistent with our analysis in the spectral information that for warm stars the B band does not behave as well as the R band while it supersedes R band for K-type stars. For metal-poor stars, it is dangerous to use B band or R band alone to derive stellar labels for F- and G-type stars due to the lack of \teff-indicators in B band and the lack of \mh-indicators in the R band. As a suggestion, targeting more objects that are observed in LRS or combining with other spectroscopic, photometric, astrometric and asteroseismic surveys may be beneficial for the MRS survey.
\item We estimated \textit{precision limits} for the abundances of $\sim$ 90 elements with gradient spectra. Taking advantages of the medium-resolution ($R\sim7,500$), we expect abundances of at least 17 elements to be measured precisely in the MRS spectra.
\item We also tested the influence on stellar labels introduced by erroneous RV. The simulated results show that within the precision of RV for MRS currently, we do not see a significant increase in the scatter. Note that the reported RV precision is mostly based on cool stars.
\item We did not take into account the binary and multiple systems, but we do see the need for identification of binary systems before deriving stellar labels using MRS spectra.
\end{enumerate}

\appendix
\section{The Percentage of Variance Explained (PVE)}
\label{sec:ve}
This section introduces the concept of the \textit{percentage of variance explaied} (PVE).
Assuming we have a mock data set containing features $x_i$ and observations $y_i$.
An \textit{ideal} regression model whose model complexity matches the data, is then fitted to the mock data.
Assuming we have $N$ observations, we can calculate the mean and variance of the observed data $y$ with 
\begin{equation}
\mu=\frac{1}{N}\sum_i^N y_i
\end{equation}
and 
\begin{equation}
s^2=\frac{1}{N}\sum_i^N (y_i-\mu)^2 .
\end{equation}
Fitting with an \textit{ideal} regression model to the data, we are able to evaluate the variance of the residuals via
\begin{equation}
s_{res}^2=\frac{1}{N}\sum_i^N (y_i-y_{mod,i})^2 .
\end{equation}
The PVE is then evaluated with
\begin{equation}
PVE = 1-\frac{s_{res}^2}{s^2}.
\end{equation}
By definition, it approaches 1 when the data contains information of feature $x$ without noise ($\mathrm{S/N}\rightarrow \infty$) and modeled properly, and it approaches 0 when information is overwhelmed by noise in data ($\mathrm{S/N}\rightarrow 0$) .
Therefore, we can use PVE to indicate the \textit{information content} of signals in noisy data.
For a systematic introduction of these concepts we refer to \citet{hastie2009}

\begin{figure}[!htbp]
    \centering
    \includegraphics[width=15cm, angle=0]{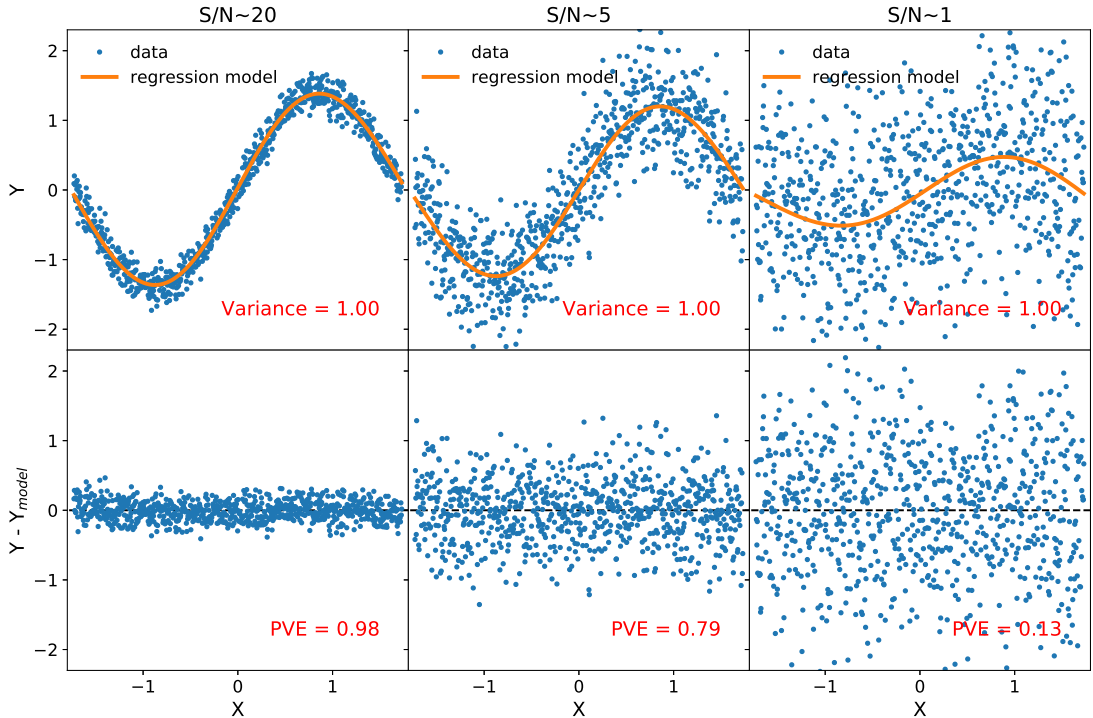}
    \caption{The demonstration of PVE values for high S/N, modest S/N and low S/N data are shown in left, middle and right panels, respectively. The upper panels show the standardized mock data and the regression model, and the lower panels show the residuals.} 
    \label{fig:pve}
\end{figure}

We show a demo to explain it a bit more.
We generate mock data with $y=\sin{x}+\epsilon$, where the Gaussian random noise term $\epsilon\sim\mathcal{N}(0,0.01)$, $\mathcal{N}(0,0.16)$ and $\mathcal{N}(0,4)$, which corresponds to S/N ratios of 20,5 and 1, respectively.
\textit{Ideal} models are fitted to the three data sets and shown in Figure~\ref{fig:pve}.
Note that both $x$ and $y$ are standardized to have a zero mean and a unity variance for visualization.
The upper panels show the data and regression models while the lower panels show the residuals.
In these three cases, including high S/N, modest S/N, and low S/N, we see that the $PVE=0.97$, $0.79$ and $0.13$, respectively.
The PVE-indicated \textit{information content} of signals are, therefore, consistent with our understanding of data.

\normalem
\begin{acknowledgements}

The authors thank the referee for providing many useful suggestions.

This work is supported by National Key R\&D Program of China No. 2019YFA0405501.
CL thanks the National Natural Science Foundation of China (NSFC) with grant No. 11835057. 


\end{acknowledgements}
  
\bibliographystyle{raa}
\bibliography{bibtex}

\end{document}